\documentclass{article}
\usepackage{geometry}
\usepackage{graphicx}
\usepackage{dcolumn}
\usepackage{bm}
\usepackage[numbers]{natbib}
\usepackage{slashed}
\usepackage{hyperref}
\usepackage{contour}
\usepackage{lmodern}
\usepackage{ulem}
\usepackage{amsmath}
\usepackage{mathrsfs}
\usepackage{multirow}
\usepackage{xcolor}
\usepackage{booktabs}
\usepackage{graphicx}
\usepackage[caption=false]{subfig}
\usepackage{feynmf}
\usepackage{tcolorbox}
\usepackage{upgreek}
\usepackage{float}
\usepackage{placeins}
\usepackage[switch*]{lineno}

\newcommand{\BR}[2]{\textrm{BR}(#1\rightarrow#2)}
\newcommand{\cW}{\tilde{c}_{W}}
\newcommand{\cZ}{\tilde{c}_{Z}}
\newcommand{\cH}{\tilde{c}_{H}}

\newcommand{\PNWA}{\textrm{P}_\textrm{NWA}(M_Q,\vec{c})}

\begin{document}

\title{Non-resonant Diagrams for Single Production of Top and Bottom Partners}



\author{Avik Roy\footnote{avroy@illinois.edu}\\
{\small{
National Center for Supercomputing Applications, University of Illinois at Urbana-Champaign
}}\\
Timothy Andeen\footnote{tandeen@utexas.edu}\\
{\small{Center for Particles and Fields, Department of Physics, The University of Texas at Austin
}}
}

\maketitle


\begin{abstract}
The search for Top and Bottom Partners is a major focus of analyses at both the ATLAS and CMS experiments. 
When singly produced, these vector-like partners of the
Standard Model third generation quarks retain a sizeable cross-section that makes them attractive search candidates in their respective topologies.
While most efforts have concentrated on the resonant mode for single production of these hypothetical particles, the most dominant mode at narrow widths, recent studies have revealed a wide and rich phenomenology involving the non-resonant diagrams.
In this letter we categorically investigate the impact of the non-resonant diagrams on different Top and Bottom partner single production topologies and their impact on the inclusive cross-section estimation.
We also propose a parameterization for calculating the the correction factor to the single vector-like quark production cross-section when such diagrams are included. 
\end{abstract}




\section{Introduction}\label{sec:intro}
Vector-like Quarks (VLQs) are predicted by a wide variety of Beyond Standard Model (BSM) theories~\cite{new-fermion, comp-higgs, comp-higgs-2, comp-higgs-3, littlest-higgs, extra-dim, extra-dim-2, extra-dim-bounds, VL-fermion, TP-LHC} at the TeV scale. With the possible extension of the Standard Model (SM) with a fourth, chiral generation of quarks heavily constrained by precision Higgs measurements~\cite{higgs-gen4-1,higgs-gen4-2}, VLQs remain a plausible choice for new quarks to extend the SM with additional quarks. Often inspired by the potential solution to the Higgs mass naturalness problem~\footnote{For a brief review of the naturalness problem for the Higgs boson mass, see Ref.~
\cite{higgs-nat-rev}.}, the cut-off scale for new physics is anticipated at the TeV scale when the fine tuning scale for the bare Higgs mass is constrained to $\mathcal{O}(1-10)$~\cite{tasi-1,tasi-2}. Hence, these hypothetical heavy quarks are prominent search candidates in the post-Higgs discovery era at CERN's Large Hadron Collider (LHC). 

In most representations, VLQs are $SU(3)$ color-triplets with strong interactions identical to the SM quarks while their left- and right-chiral representations remain same within the electroweak (EW) symmetry group. The usual four VLQ species are denoted as $X_{+\frac{5}{3}}, T_{+\frac{2}{3}}, B_{-\frac{1}{3}}$ and $Y_{-\frac{4}{3}}$ where the subscript indicates the electric charge of the corresponding particle. Their EW representation identify $(T)$ or $(B)$ singlets, $(X,T)$, $(T,B)$, or $(B,Y)$ doublets and $(X,T,B)$ or $(T, B, Y)$ triplets. In most representations, they couple to the Standard Model quarks via exchange of charged $(W^+, W^-)$ or neutral $(Z, H)$ bosons. The interaction of the VLQs with the SM quarks can be summarized with the following simplified Lagrangian~\cite{Wulzer},

\begin{equation}\label{SimpLag}
\mathcal{L} = \sum_{\zeta, q, Q} \left[ \frac{g_w}{2} \sum_{V} c_{\zeta,V}^{Qq}\bar{Q}_\zeta \slashed{V} q_\zeta + c_{\zeta, H}^{Qq} H \bar{Q}_{\zeta '} q_\zeta \right] + \mathrm{h.c.}
\end{equation}
where $Q$ represents the usual VLQs, $\zeta$ and $\zeta '$ represent alternate chiralities and $q$ represents a SM quark of up or down type. Some of these couplings may be constrained by the conservation of certain quantum numbers. For example, the $+\frac{5}{3}$ charged partner $X$ can only couple with SM up-type quarks by the exchange of a $W$ boson. The usual mass hierarchy of VLQs suggests they interact predominantly with the third generation of the SM quarks~\cite{okada,aguilar-2,exot-6}. Hence, VLQs are often dubbed Top or Bottom Partners and their interaction with lighter generations set to zero in the simplified representation of Eqn.~\ref{SimpLag}. In this paper, we will generally focus on Top and Bottom partners, denoted by $T$ and $B$ respectively, and unless otherwise stated, the term VLQ will be used interchangeably with $T$ and $B$.

\begin{figure*}[!ht]
\centering
\subfloat[]{
\begin{fmfgraph*}(100,100)
   \fmfleft{i1,i2}
   \fmfright{o1,o2}
   \fmf{gluon,tension=1}{i1,v1} 
   \fmf{gluon,tension=1}{i2,v1}
   \fmf{gluon,tension=1}{v1,v2}
   \fmf{fermion,tension=0.6,label=$Q$,label.side=right}{v2,o1}
   \fmf{fermion,tension=0.6,label=$\bar{Q}$}{o2,v2}
\end{fmfgraph*}
\label{fig:feyn-PP}
}
\subfloat[]{
\begin{fmfgraph*}(100,100)
   \fmfleft{i1,i2}
   \fmfright{o1,o2,o3,o4}
   \fmf{gluon,tension=3}{i1,v1} 
   \fmf{fermion,tension=2}{o1,v1}
   \fmf{fermion,tension=3}{v1,v2}
   \fmf{fermion,tension=3}{i2,v3}
   \fmf{fermion,tension=2}{v3,o4}
   \fmf{boson,tension=3}{v3,v2}
   \fmf{fermion,tension=1,label=$Q$}{v2,v4}
   \fmf{boson,tension=0.2}{v4,o3}
   \fmf{dashes,tension=0.2}{v4,o3}
   \fmf{fermion,tension=0.5}{v4,o2}
\end{fmfgraph*}
\label{fig:feyn-SP-s}
}
\subfloat[]{
\begin{fmfgraph*}(100,100)
   \fmfleft{i1,i2}
   \fmfright{o1,o2,o3,o4}
   \fmf{gluon,tension=3}{i1,v1} 
   \fmf{fermion,tension=2}{o1,v1}
   \fmf{fermion,tension=1.5}{v1,v2}
   \fmf{fermion,tension=2.8}{i2,v3}
   \fmf{fermion,tension=2}{v3,o4}
   \fmf{boson,tension=2}{v3,v4}
   \fmf{fermion,tension=2,label=$Q$}{v2,v4}
   \fmf{boson,tension=0.2}{v2,o2}
   \fmf{dashes,tension=0.2}{v2,o2}
   \fmf{fermion,tension=0.5}{v4,o3}
\end{fmfgraph*}
\label{fig:feyn-SP-t}
}
\label{fig:feyns-1}
\caption{Dominant contributing diagrams for \protect\subref{fig:feyn-PP} pair production of VLQs, \protect\subref{fig:feyn-SP-s} single production of VLQs in $s$-channel topology, and \protect\subref{fig:feyn-SP-t} single production of VLQs in $t$-channel topology }
\label{fig:PPSP-diag}
\end{figure*}
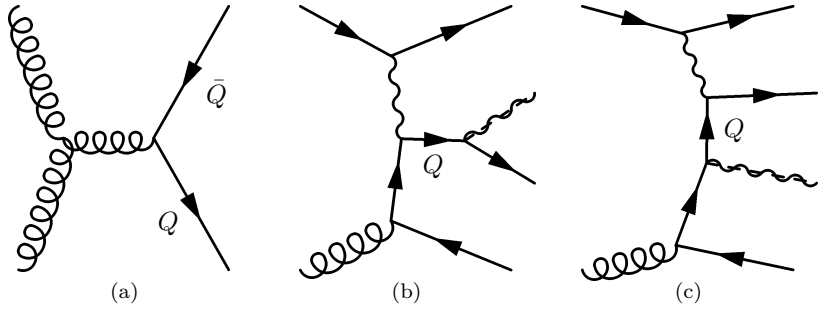

ATLAS and CMS have performed extensive searches for both pair-produced and singly produced Top and Bottom partners. During Run I of the LHC, at a center of mass energy of 8 TeV, searches for pair production of VLQs were more popular compared to searches for singly produced VLQs~\cite{ATLAS-run1-PP-1, ATLAS-run1-PP-2, ATLAS-run1-PP-3, CMS-run1-PP-1, CMS-run1-PP-2, CMS-run1-PP-3}. At the lower center of mass energy with the focus of analysis typically confined to VLQ masses less than 1 TeV, the production cross-section of pairs of VLQ is dominant~\cite{Aguilar}. Additionally, the strong force mediated pair production (Figure \ref{fig:feyn-PP}) of VLQs becomes a model-independent process and hence the process cross-section only depends on the VLQ mass and not on the species nor its representation. 
No significant excess was seen in data and exclusion limits on  VLQ mass were reported in the range of approximately \mbox{600--1000 GeV}. These results have served as the benchmark for the Run II searches at 13 TeV center of mass energy. The search effort for single $T$ or $B$ also reported very similar limits in Run I~\cite{ATLAS-run1-PP-3, ATLAS-TWb-run1}.

\begin{fmffile}{VLQ-Feyns}
\end{fmffile}


The Run II ATLAS searches using $36.1 \mathrm{fb}^{-1}$ of data from 2015-16 for the pair production of  Top and Bottom Partners~\cite{ATLAS-run2-PP-1, ATLAS-run2-PP-2, ATLAS-run2-PP-4,  ATLAS-run2-PP-5,  ATLAS-run2-PP-6} and their statistical combination~\cite{ATLAS-PPcomb} set the strongest current limits on the VLQ masses, excluding $T (B)$ masses up to 1.31~(1.03)~TeV for any combination of VLQ branching ratios. Complementary pair production analyses from CMS~\cite{CMS-run2-PP-1, CMS-run2-PP-2, CMS-run2-PP-3} have also set limits in the range of $\mathcal{O}(1 \textrm{ TeV})$ for Top and Bottom Partners. 

With the region of sensitivity pushed to higher masses, there has been a significant increase in the number of searches for singly produced VLQs~\cite{ATLAS-TZt-run2, ATLAS-TWb-run2, ATLAS-monotop, CMS-TZt-run2,  CMS-XWt-run2, CMS-BHb-run2, CMS-fullhad-run2} during Run II. This is partly because, depending on how strongly VLQs couple with SM bosons and quarks, single production processes can have a larger cross-section at the range of masses that Run II searches have been focusing on~\citep{Aguilar}. However, unlike pair production, the production of single VLQs is dominated by electroweak processes (Figures \ref{fig:feyn-SP-s},\ref{fig:feyn-SP-t}). Both kinematics and cross-section of single VLQ processes depend on the representation and the choice of couplings that determine the relative strength of these particles interacting with SM quarks and vector/Higgs bosons. Beyond the usual scope of VLQs at narrow width at small couplings, recent studies have explored VLQ phenomenology at finite width and at next-to-leading-order (NLO)~\cite{Panizzi-LW-2,Panizzi-LW,Panizzi-LW-3,Panizzi-NLO,interpretation}. These studies suggest that extrapolating single VLQ search results performed under the narrow width approximation to large width by parametric recasting alone is neither trivial nor sufficient. A consistent approach for modeling finite width effects requires taking into account the effects of finite width in parton-level simulations. 

This paper aims at addressing the relative contribution of resonant (the $s$-channel in Figure~\ref{fig:feyn-SP-s}) and nonresonant (the $t$-channel in Figure~\ref{fig:feyn-SP-t}) topologies in single VLQ processes at the LHC. In the following section, we investigate the dominant $s$- and $t$-channel topologies for different $VQAq$ process where $VQAq$ is short-hand notation for the single production of the VLQ $Q$ being mediated by the vector boson $V$ where the final state includes the boson $A$ and SM quark $q$. We use a four flavor scheme for parton distributions in the colliding protons. This, along with the assumption of narrow width for all SM bosons and fermions, allows us to factorize the SM particle decays such that they do not impact the parton level kinematic distributions.

\section{$t$-channel topology beyond narrow width}\label{sec:topo}

At narrow width, when the couplings are small enough to allow $\frac{\Gamma_Q}{M_Q} \rightarrow 0$, the $t$-channel diagram has a negligible contribution. However, this contribution can become significant for finite VLQ decay widths. This can be understood by examining the gluon splitting in $s$- and $t$-channel diagrams. To illustrate this point, we show the corresponding Feynman diagrams for $WTZt$ and $ZTZt$ processes in Figure~\ref{fig:feyn-WZTZt}. As for the $WTZt$ process, the matrix element for the $s$-channel diagram receives a large contribution from the gluon splitting to a pair of $b$ quarks while the corresponding $t$-channel diagram is further suppressed by the kinematically less favorable splitting of gluon to a pair of $t$ quarks. In accordance with the findings in Ref.~\cite{Panizzi-LW-3}, the $t$-channel consistently remains marginal for all three of the $W$-mediated produced single-$T$ processes, namely $WTWb, WTZt,$ and $WTHt$. 

\begin{fmffile}{WT-ZT-Feyns}
\begin{figure*}[!ht]
\centering
\subfloat[]{
\begin{fmfgraph*}(150,150)
   \fmfleft{i1,i2}
   \fmfright{o1,o2,o3,o4}
   \fmf{gluon,tension=3,label=$g$}{i1,v1} 
   \fmf{fermion,tension=2,label=$\bar{b}/\color{red}{\bar{t}}$}{o1,v1}
   \fmf{fermion,tension=4.5, label=$b/\color{red}{t}$}{v1,v2}
   \fmf{fermion,tension=3,label=$q$}{i2,v3}
   \fmf{fermion,tension=2,label=$q'/\color{red}{q}$}{v3,o4}
   \fmf{boson,tension=3,label=$W/\color{red}{Z}$}{v3,v2}
   \fmf{fermion,tension=1,label=$T$}{v2,v4}
   \fmf{boson,tension=0.2,label=$Z$}{v4,o3}
   \fmf{fermion,tension=0.5,label=$t$}{v4,o2}
\end{fmfgraph*}
\label{fig:feyn-WZTZt-s}
}
\subfloat[]{
\begin{fmfgraph*}(150,150)
   \fmfleft{i1,i2}
   \fmfright{o1,o2,o3,o4}
   \fmf{gluon,tension=3,label=$g$}{i1,v1}
   \fmf{fermion,tension=2,label=$\bar{t}$}{o1,v1}
   \fmf{fermion,tension=2.3,label=$t$}{v1,v2}
   \fmf{fermion,tension=2.8,label=$q$}{i2,v3}
   \fmf{fermion,tension=2,label=$q'/\color{red}{q}$}{v3,o4}
   \fmf{boson,tension=2,label=$W/\color{red}{Z}$}{v3,v4}
   \fmf{fermion,tension=2,label=$T$}{v2,v4}
   \fmf{boson,tension=0.8,label=$Z$}{v2,o2}
   \fmf{fermion,tension=0.5,label=$t$}{v4,o3}
\end{fmfgraph*}
\label{fig:feyn-WZTZt-t}
}
\label{fig:feyns-2}
\caption{\protect\subref{fig:feyn-WZTZt-s} $s$-channel and \protect\subref{fig:feyn-WZTZt-t} $t$-channel diagrams for $WTZt$ and {\color{red}{{$ZTZt$}}} processes.}
\label{fig:feyn-WZTZt}
\end{figure*}
\end{fmffile}

However, as we look at the $ZTZt$ process diagrams, we note with interest that both $s$- and $t$-channel diagrams actually receive a contribution from the same gluon splitting function, i.e. gluon splitting to a pair of top quarks. As we consider larger decay widths for the Top Partner, the relative contribution of the $t$-channel diagram can make an observable difference, especially when the interference between the two channels is considered. 

In order to investigate the impact of $t$-channel diagrams on single-$T$ and single-$B$ processes, we simulate these processes in \textsc{MadGraph\_aMC@NLO}~\cite{MG-1,MG-2} using the VLQ UFO model inspired by the parameterization presented in~\cite{Fuks}. For each of the $VQAq$ processes with $Q \in \{T,B\}$, we generated 20,000 events for $M_Q$ in the range of \mbox{1000--2200 GeV} in steps of \mbox{200 GeV} while the relative decay widths are varied between $0.01$ and $0.5$. The $\kappa, \hat{\kappa}, \tilde{\kappa}$ parameters introduced in~\cite{Fuks} have been calculated from the rescaled couplings $\cW,\cZ,\cH$ introduced in Ref.~\cite{interpretation} from a one-to-one correspondence with the tree-level couplings in equation (\ref{SimpLag}) while requiring $\cW = \cZ = \cH$, which would imply almost equal branching ratios in the three decay channels for $T$ and $B$ quarks at the narrow width limit. 

\begin{figure*}[!ht]
\centering
  \subfloat[]{
  \includegraphics[width=0.25\textwidth]{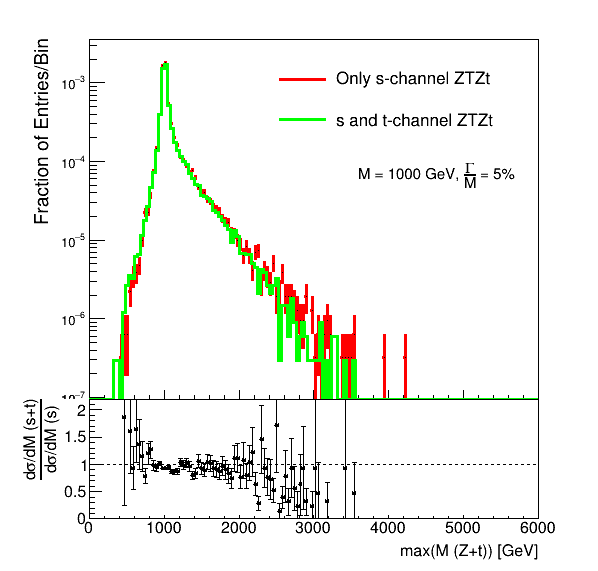}
  \label{fig:mZT-10-5}
  }
  \subfloat[]{
  \includegraphics[width=0.25\textwidth]{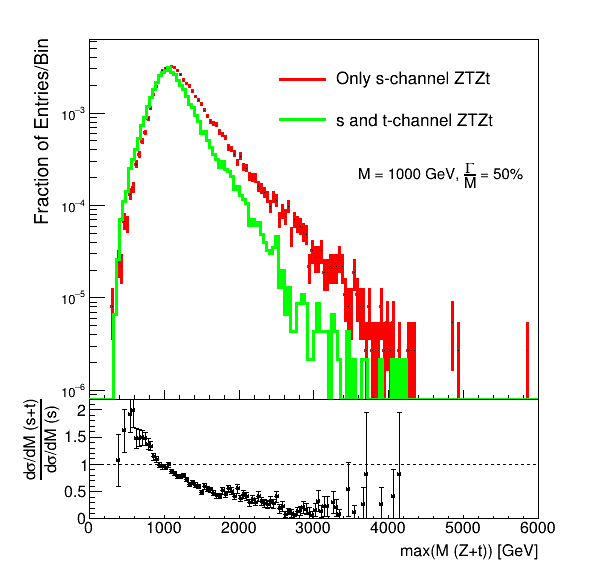}
  \label{fig:mZT-10-50}
  }
  \subfloat[]{
  \includegraphics[width=0.25\textwidth]{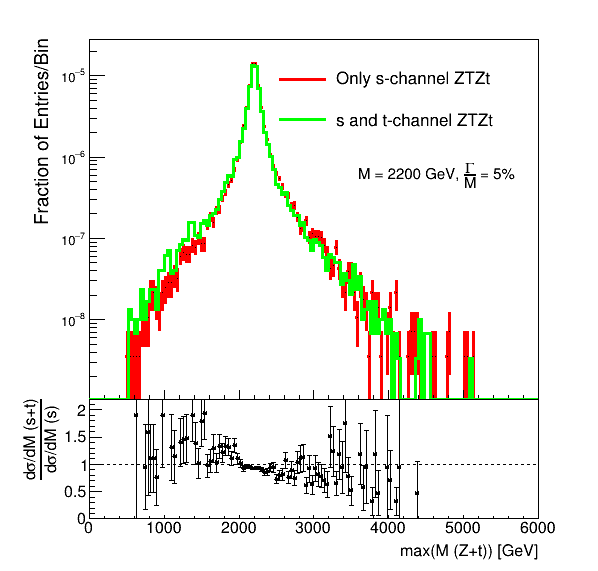}
  \label{fig:mZT-22-5}
  }
  \subfloat[]{
  \includegraphics[width=0.25\textwidth]{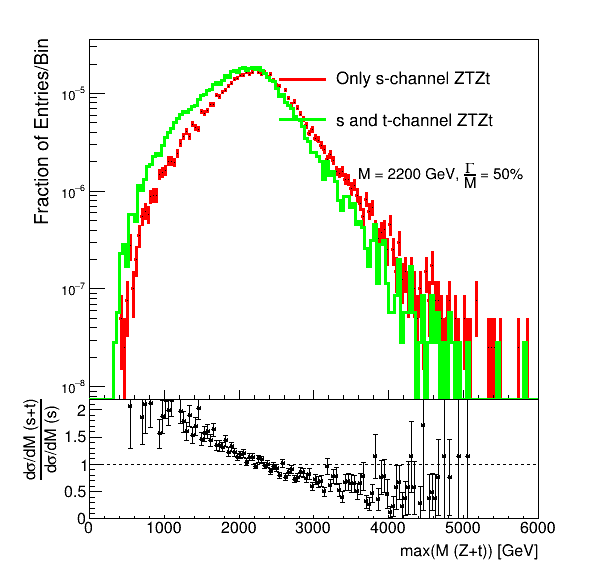}
  \label{fig:mZT-22-50}
  }
  \caption[]{Distribution of the leading invariant mass of the $Z$ boson and the $t$ quark for
  \protect\subref{fig:mZT-10-5} $M_T = 1.0$~TeV and $\frac{\Gamma}{M} = 0.05$,
  \protect\subref{fig:mZT-10-50} $M_T = 1.0$~TeV and $\frac{\Gamma}{M} = 0.5$,
  \protect\subref{fig:mZT-22-5} $M_T = 2.2$~TeV and $\frac{\Gamma}{M} = 0.05$, and
  \protect\subref{fig:mZT-22-50} $M_T = 2.2$~TeV and $\frac{\Gamma}{M} = 0.5$.
} 
  \label{fig:mZT-10-22}
\end{figure*}

Figure~\ref{fig:mZT-10-22} shows the distribution of the leading invariant mass of the $Z$ boson and the $t$ quark in the final state of the $ZTZt$ processes for $M_T = 1.0$ and $2.2$~TeV at relative widths of $5\%$ (narrow width) and $50\%$ (large width). As expected, the non-resonant contribution at narrow width (Figures~\ref{fig:mZT-10-5}~and~\ref{fig:mZT-22-5}) is rather minimal and shows up only at the tails of the distributions where the differential cross-section is already suppressed by a few orders of magnitude. However, as larger widths are probed (Figures~\ref{fig:mZT-10-50}~and~\ref{fig:mZT-22-50}), the effect of interference between the $s$- and $t$-channel diagrams become more prominent, causing a clear shift in the differential spectrum.

\begin{table}[!ht]
    \centering
    \begin{tabular}{|c | c|}
    \hline
    $s$- and $t$-channel & \multirow{2}{*}{$VQAq$ modes} \\
      gluon splittings & \\
    \hline
    $b\bar{b}$, $b\bar{b}$ & $WTWb, ZBZb, ZBHb$ \\
    $b\bar{b}$, $t\bar{t}$ & $WTHt, WTZt, ZBWt$ \\
    $t\bar{t}$, $b\bar{b}$ & $WBHb, ZTWb, WBZb$ \\
    $t\bar{t}$, $t\bar{t}$ & $ZTHt, ZTZt, WBWt$ \\
    \hline
    \end{tabular}
    \caption{Gluon splitting modes in $s$- and $t$-channel diagrams for different $VQAq$ processes for single-$T$ and single-$B$}
    \label{tab:gluon_split}
\end{table}

\begin{figure*}[!ht]
\centering
  \subfloat[]{
  \includegraphics[width=0.25\textwidth]{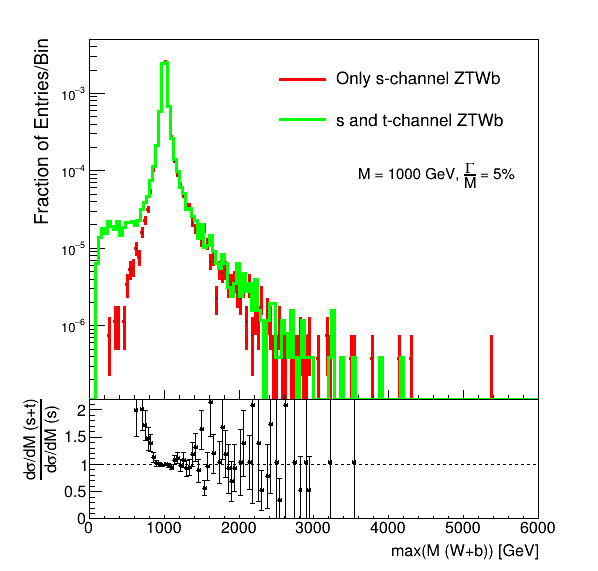}
  \label{fig:mWb-10-5}
  }
  \subfloat[]{
  \includegraphics[width=0.25\textwidth]{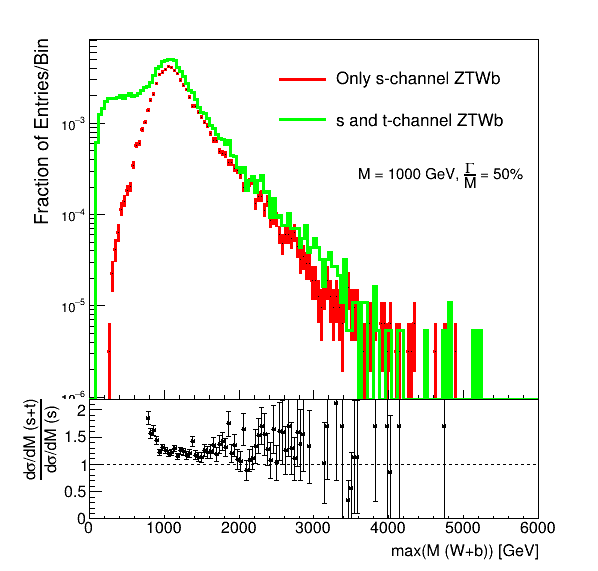}
  \label{fig:mWb-10-50}
  }
  \subfloat[]{
  \includegraphics[width=0.25\textwidth]{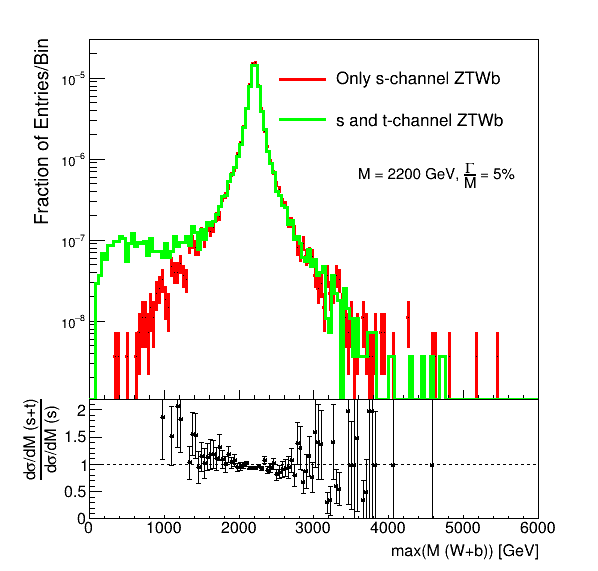}
  \label{fig:mWb-22-5}
  }
  \subfloat[]{
  \includegraphics[width=0.25\textwidth]{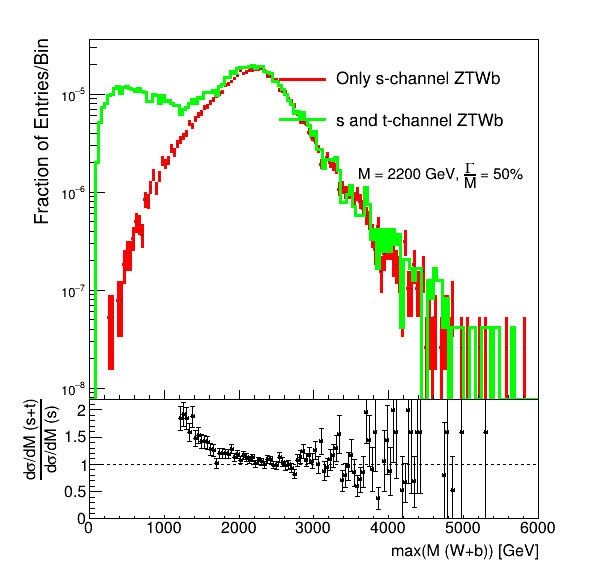}
  \label{fig:mWb-22-50}
  } \\
  \subfloat[]{
  \includegraphics[width=0.25\textwidth]{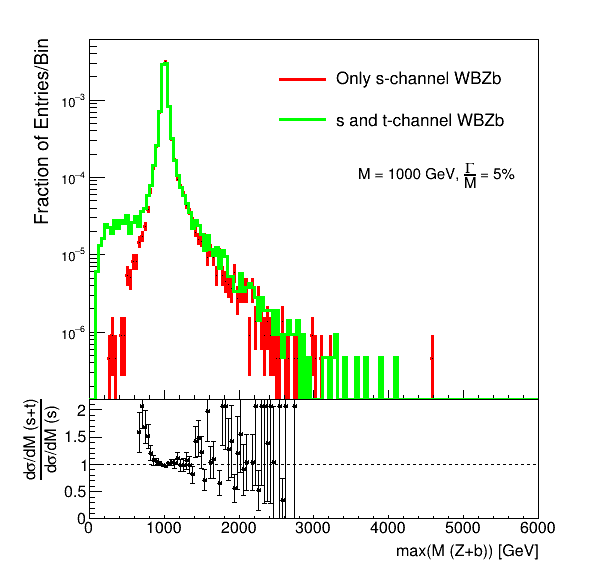}
  \label{fig:mZb-10-5}
  }
  \subfloat[]{
  \includegraphics[width=0.25\textwidth]{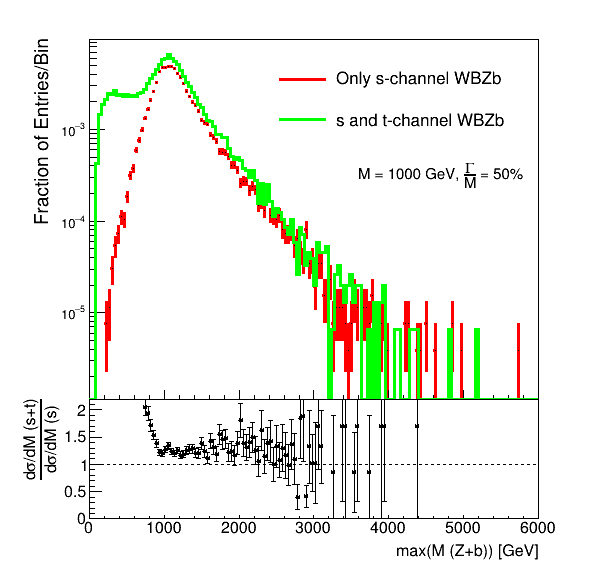}
  \label{fig:mZb-10-50}
  }
  \subfloat[]{
  \includegraphics[width=0.25\textwidth]{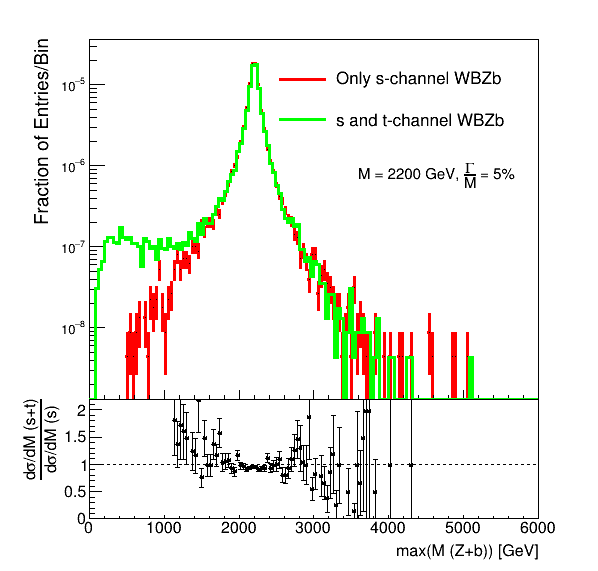}
  \label{fig:mZb-22-5}
  }
  \subfloat[]{
  \includegraphics[width=0.25\textwidth]{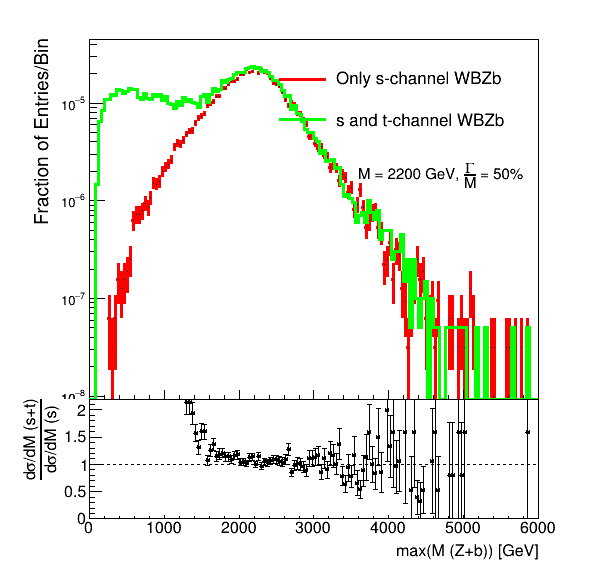}
  \label{fig:mZb-22-50}
  }
  \caption[]{Distribution of the leading $W+b$ (top row, for $ZTWb$) and {\color{red}{$Z+b$ (bottom row, for $WBZb$)}} inavraint masses for
  \protect\subref{fig:mWb-10-5},
  {\color{red}{\protect\subref{fig:mZb-10-5}}} 
  $M_{T/\color{red}{B}} = 1.0$~TeV and $\frac{\Gamma}{M} = 0.05$,
  \protect\subref{fig:mWb-10-50},
  {\color{red}{\protect\subref{fig:mZb-10-50}}}
  $M_{T/\color{red}{B}} = 1.0$~TeV and $\frac{\Gamma}{M} = 0.5$,
  \protect\subref{fig:mWb-22-5},
  {\color{red}{\protect\subref{fig:mZb-22-50}}}
  $M_{T/\color{red}{B}} = 2.2$~TeV and $\frac{\Gamma}{M} = 0.05$, 
  \protect\subref{fig:mWb-22-50},
  {\color{red}{\protect\subref{fig:mZb-22-50}}}
  $M_{T/\color{red}{B}} = 2.2$~TeV and $\frac{\Gamma}{M} = 0.5$.
} 
  \label{fig:mWbZb-10-22}
\end{figure*}

As the shapes of kinematic distributions associated with $T$ or $B$ quark decay can be significantly affected by the choice of contribution from non-resonant diagrams with similar gluon splitting diagrams, it is even more interesting to investigate similar impacts when the $t$-channel diagram receives a contribution from gluon splitting to $b\bar{b}$ while the resonant contribution of the $s$-channel diagram receives contribution from the $t$-channel diagram. The gluon splitting modes of different $VQAq$ processes are listed in Table~\ref{tab:gluon_split}. Both $ZTWb$ and $WBZb$ modes require the presence of a gluon splitting to $b\bar{b}$ in the non-resonant diagram and a $t\bar{t}$ splitting for the resonant diagram in $s$-channel. The differential cross-section with respect to the leading invariant masses of $W+b$ and $Z+b$ for these processes are shown in Figure~\ref{fig:mWbZb-10-22}. The usual assumption of a negligible impact of $t$-channel still holds for these processes (Figures~\ref{fig:mWb-10-5}~,~\ref{fig:mWb-22-5}~,~\ref{fig:mZb-10-5}~and~\ref{fig:mZb-22-5}). In this case the $t$-channel contribution is only apparent in the low mass tail. Here the differential cross-section is up to two orders of magnitude smaller than what is found near the peak of the distribution. However, at larger widths (Figures~\ref{fig:mWb-10-50}~,~\ref{fig:mWb-22-50}~,~\ref{fig:mZb-10-50}~and~\ref{fig:mZb-22-50}), these distributions receive a sizeable non-resonant contribution at the lower mass tails. It is apparent that these contributions not only change the shape of the overall distribution but also introduces a sizeable increase in the process cross-section overall. 

\begin{figure*}[!ht]
\centering
  \subfloat[]{
  \includegraphics[width=0.25\textwidth]{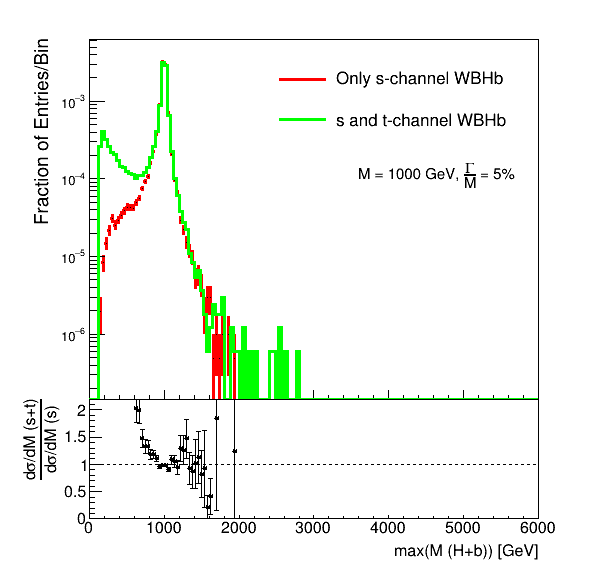}
  \label{fig:mWBHb-10-5}
  }
  \subfloat[]{
  \includegraphics[width=0.25\textwidth]{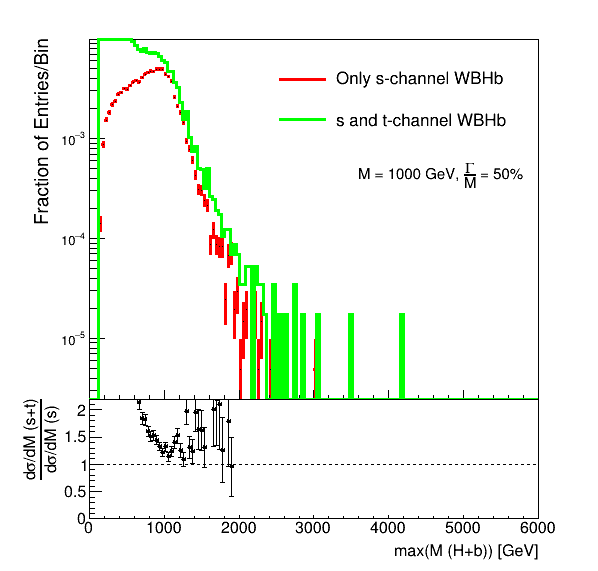}
  \label{fig:mWBHb-10-50}
  }
  \subfloat[]{
  \includegraphics[width=0.25\textwidth]{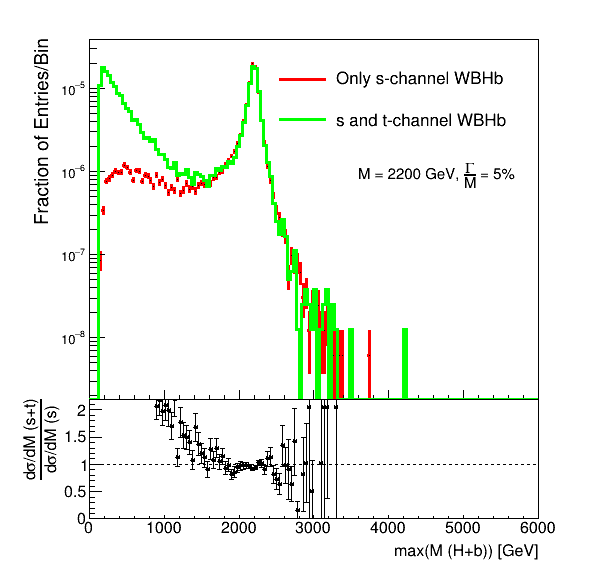}
  \label{fig:mWBHb-22-5}
  }
  \subfloat[]{
  \includegraphics[width=0.25\textwidth]{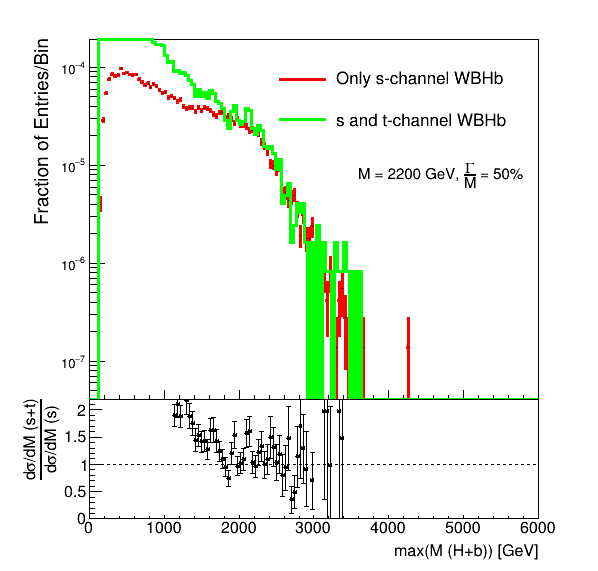}
  \label{fig:mWBHb-22-50}
  } \\
  \subfloat[]{
  \includegraphics[width=0.25\textwidth]{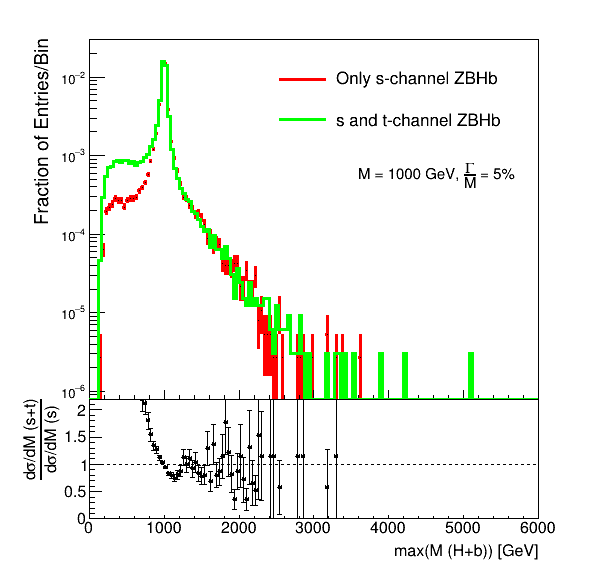}
  \label{fig:mZBHb-10-5}
  }
  \subfloat[]{
  \includegraphics[width=0.25\textwidth]{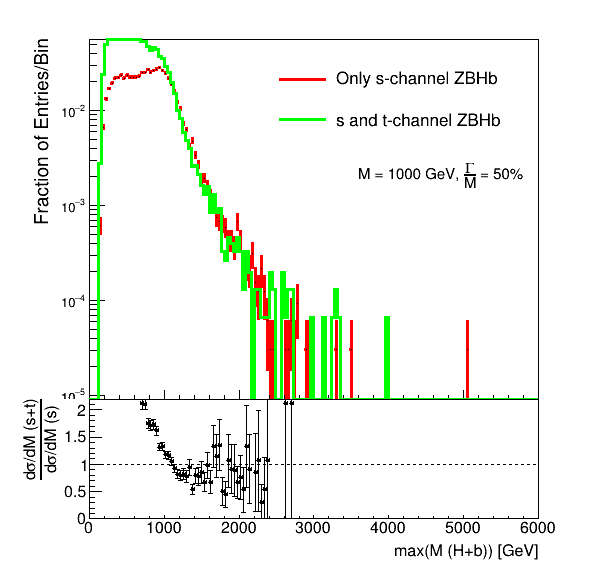}
  \label{fig:mZBHb-10-50}
  }
  \subfloat[]{
  \includegraphics[width=0.25\textwidth]{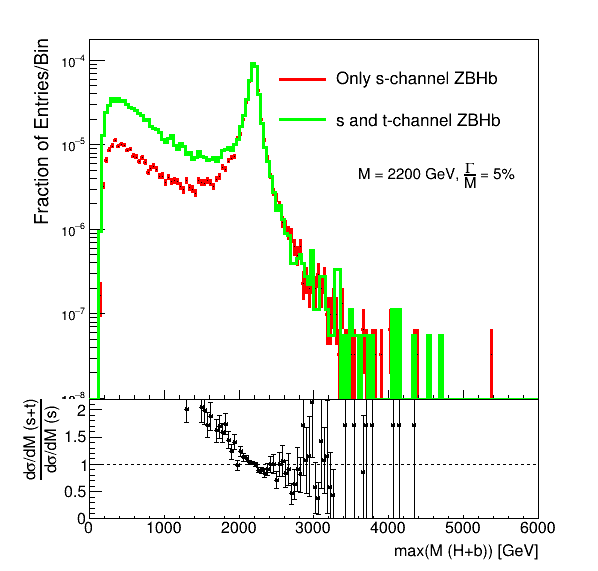}
  \label{fig:mZBHb-22-5}
  }
  \subfloat[]{
  \includegraphics[width=0.25\textwidth]{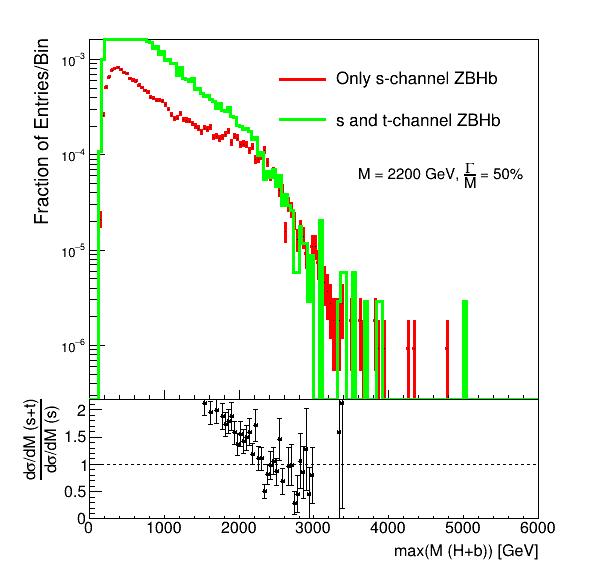}
  \label{fig:mZBHb-22-50}
  }\\
  \subfloat[]{
  \includegraphics[width=0.25\textwidth]{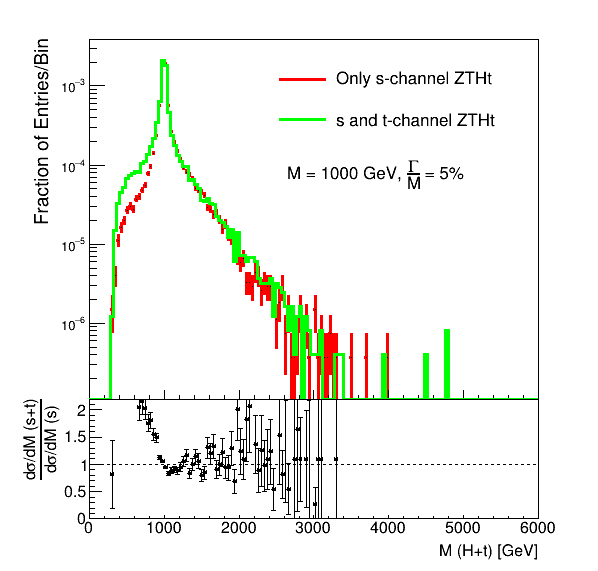}
  \label{fig:mZTHt-10-5}
  }
  \subfloat[]{
  \includegraphics[width=0.25\textwidth]{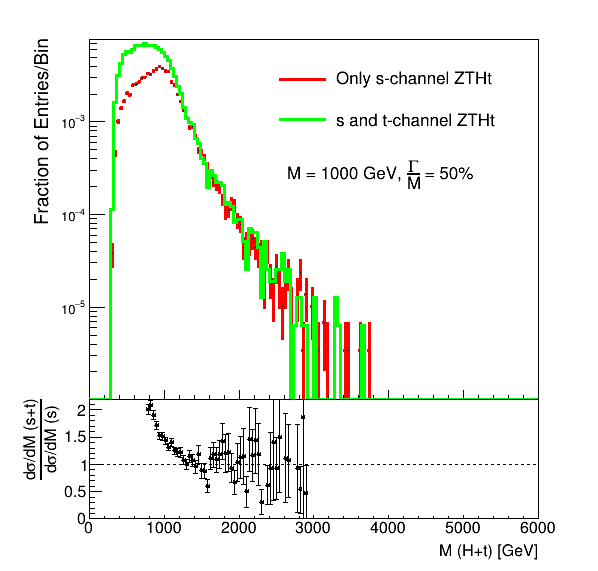}
  \label{fig:mZTHt-10-50}
  }
  \subfloat[]{
  \includegraphics[width=0.25\textwidth]{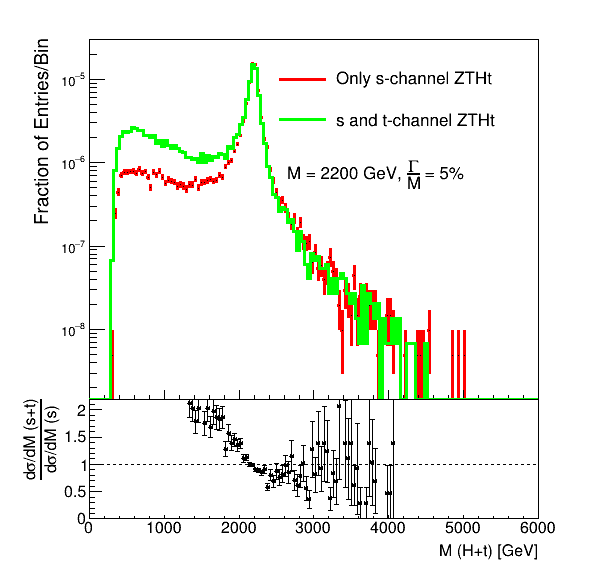}
  \label{fig:mZTHt-22-5}
  }
  \subfloat[]{
  \includegraphics[width=0.25\textwidth]{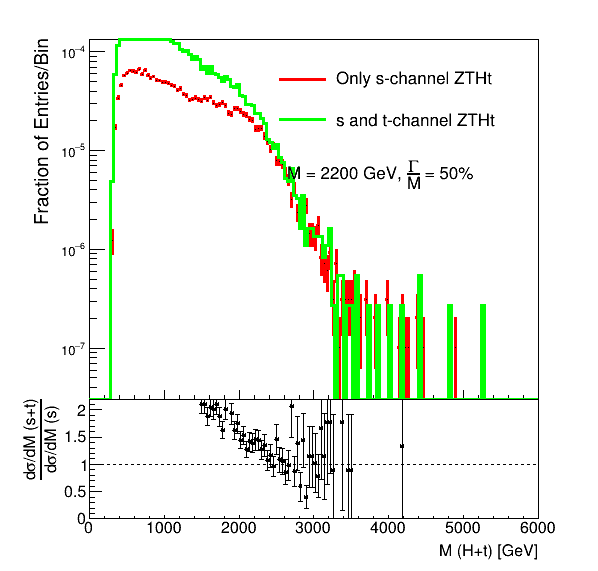}
  \label{fig:mZTHt-22-50}
  }
  \caption[]{Distribution of the leading $H+b$ (top row, for $WBHb$ and middle row, for $ZBHb$) and $H+t$ (bottom row, for $ZTHt$) inavraint masses for
  \protect\subref{fig:mWBHb-10-5},
  \protect\subref{fig:mZBHb-10-5},
  \protect\subref{fig:mZTHt-10-5},
  $M_{T/B} = 1.0$~TeV and $\frac{\Gamma}{M} = 0.05$,
  \protect\subref{fig:mWBHb-10-50},
  \protect\subref{fig:mZBHb-10-50},
  \protect\subref{fig:mZTHt-10-50},
  $M_{T/B} = 1.0$~TeV and $\frac{\Gamma}{M} = 0.50$,
  \protect\subref{fig:mWBHb-22-5},
  \protect\subref{fig:mZBHb-22-5},
  \protect\subref{fig:mZTHt-22-5},
  $M_{T/B} = 2.2$~TeV and $\frac{\Gamma}{M} = 0.05$, and
  \protect\subref{fig:mWBHb-22-50},
  \protect\subref{fig:mZBHb-22-50},
  \protect\subref{fig:mZTHt-22-50},
  $M_{T/B} = 1.0$~TeV and $\frac{\Gamma}{M} = 0.50$,
} 
  \label{fig:mHbHt-10-22}
\end{figure*}

It is now clear that the $t$-channel can make a sizeable contribution to the cross-section when the gluon splitting function has a large contribution for the non-resonant diagrams at large width. This effect will become further amplified when considering the $H$-associated final states for single VLQ processes. As described in Ref.~\cite{Panizzi-LW-3}, the resonant cross-section for single VLQ processes with Higgs associated final states includes an extra factor of $\frac{1}{s}$ that amplifies the differential cross-section at lower energies. This extra boost can be further amplified when coupled with $t$-channel diagrams that come with a comparable (or dominant) gluon splitting function. From Table~\ref{tab:gluon_split}, we can see that both $s$- and $t$-channel diagrams for $ZBHb$ ($ZTHt$) have a gluon splitting to $b\bar{b}$ ($t\bar{t}$), while the $t$-channel gluon splitting to $b\bar{b}$ for $WBHb$ has a dominant contribution compared to $s$-channel's $t\bar{t}$. In Figure~\ref{fig:mHbHt-10-22}, we show the distributions for invariant mass of the leading $H+b$  for $WBHb$ and $ZBHb$ processes as well as the distribution for leading $H+t$ invariant mass distribution for $ZTHt$. In these cases, the low mass tail can have a significant contribution even at not-so-narrow widths, and the overall production cross-section can become significantly larger than what is seen from $s$-channel production alone.

\section{Parameterized cross-section estimate for single VLQ processes}\label{sec:Ft}
At narrow width, single VLQ process cross-sections can be estimated with a factorization of couplings and branching ratios~\cite{Wulzer,interpretation}-

\begin{equation}
    \sigma_{VQAq, \mathrm{NWA}} = \tilde{c}_V^2 \times \sigma_{VQ} (\tilde{c}_V = 1) \times \BR{Q}{Aq}.
    \label{eqn:xs-nw}
\end{equation}

\noindent At larger decay widths for $T$ and $B$ quarks, the cross-section estimate of Eqn.~\ref{eqn:xs-nw} needs to be modified. In Ref.~\cite{interpretation}, a parameterized correction for the $s$-channel cross-section estimate has been formulated that modifies the factorized cross-section calculation as

\begin{equation}
    \sigma_{VQAq, s} = \frac{1}{\PNWA} \times \tilde{c}_V^2 \times \sigma_{VQ} (\tilde{c}_V = 1) \times \BR{Q}{Aq}.
    \label{eqn:xs-fw-sonly}
\end{equation}

\noindent When the $t$-channel contribution becomes significant, the estimate from Eqn.~\ref{eqn:xs-fw-sonly} needs to be further revised. Therefore we introduce a parameterized cross-section correction $F_{t,VQAq}(M_Q, \frac{\Gamma_Q}{M_Q})$, defined as-

\begin{equation}
    F_{t,VQAq}\left(M_Q, \frac{\Gamma_Q}{M_Q} \right)= \frac{\sigma_{VQAq, s+t}}{\sigma_{VQAq, s}}.
\end{equation}

\begin{figure*}[!ht]
\centering
  \subfloat[]{
  \includegraphics[width=0.25\textwidth]{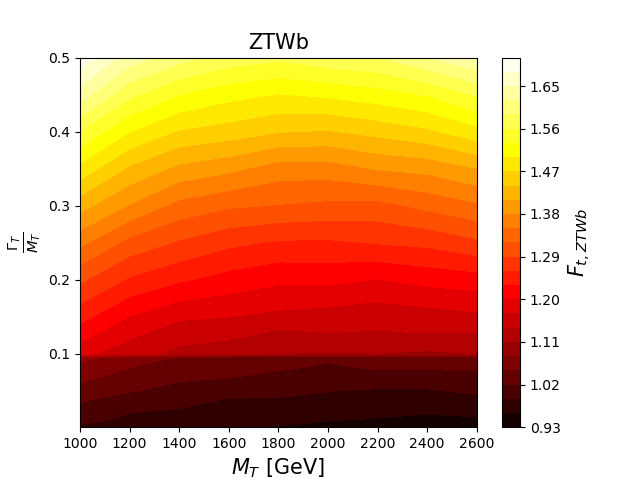}
  \label{fig:Ft-ZTWb}
  }
  \subfloat[]{
  \includegraphics[width=0.25\textwidth]{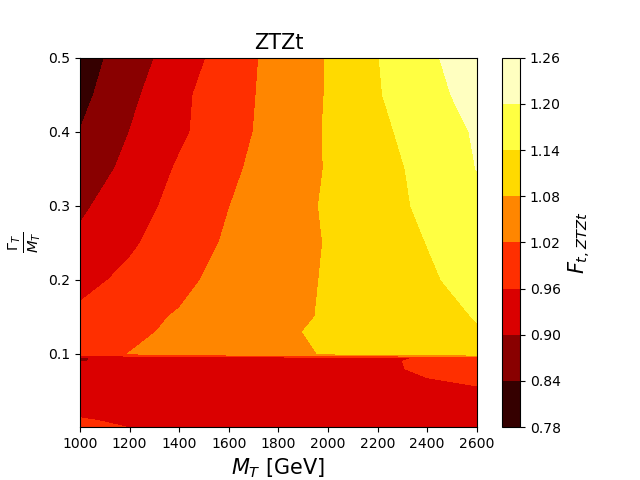}
  \label{fig:Ft-ZTZt}
  }
  \subfloat[]{
  \includegraphics[width=0.25\textwidth]{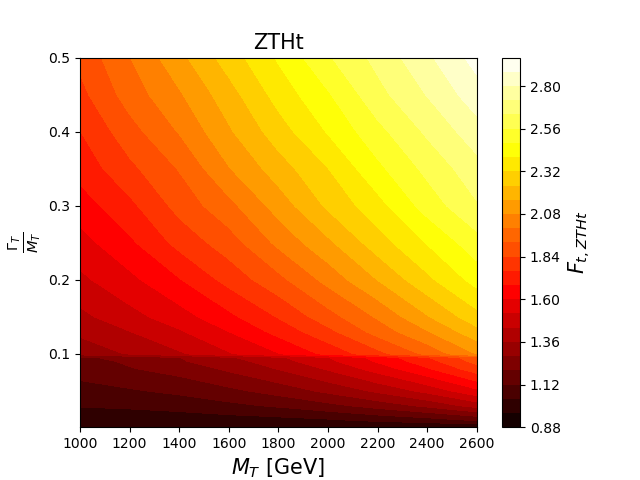}
  \label{fig:Ft-ZTHt}
  }
  \subfloat[]{
  \includegraphics[width=0.25\textwidth]{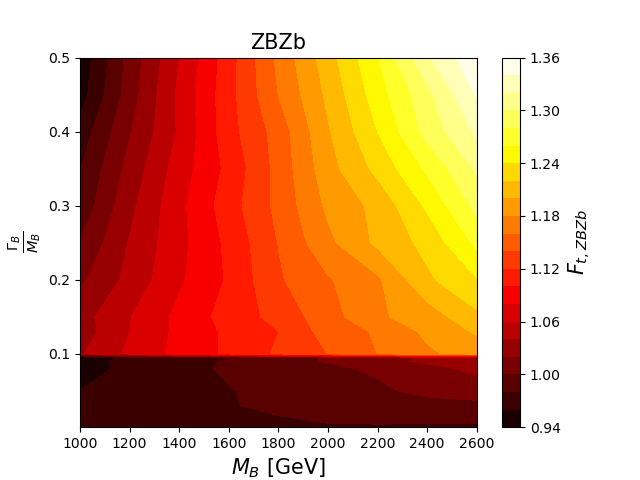}
  \label{fig:Ft-ZBZb}
  } \\
  \subfloat[]{
  \includegraphics[width=0.25\textwidth]{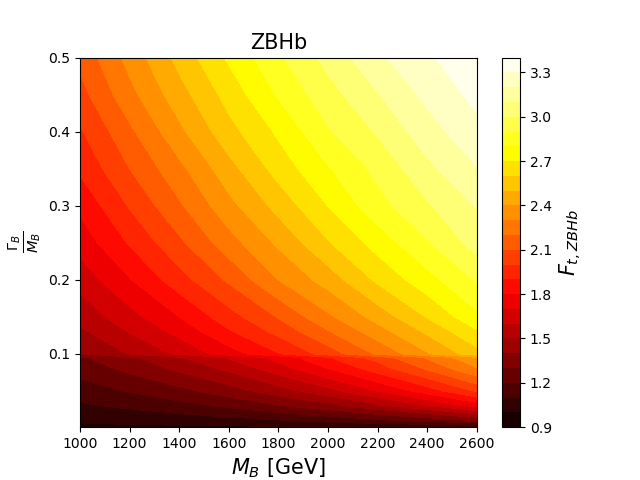}
  \label{fig:Ft-ZBHb}
  }
  \subfloat[]{
  \includegraphics[width=0.25\textwidth]{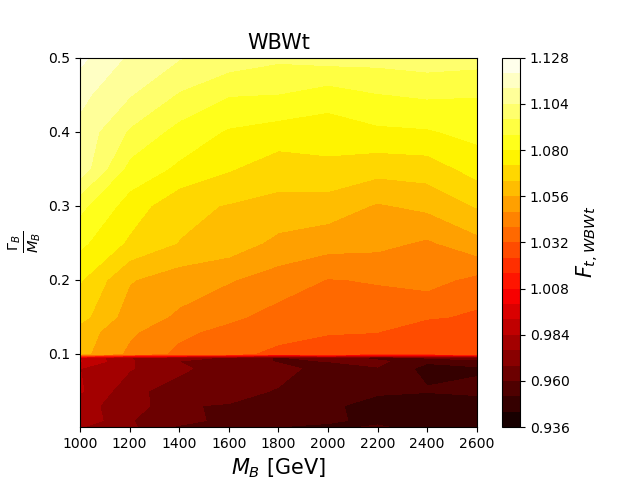}
  \label{fig:Ft-WBWt}
  }
  \subfloat[]{
  \includegraphics[width=0.25\textwidth]{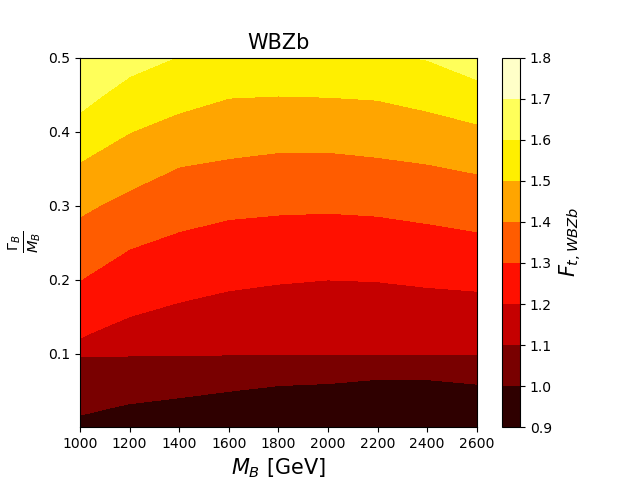}
  \label{fig:Ft-WBZb}
  }
  \subfloat[]{
  \includegraphics[width=0.25\textwidth]{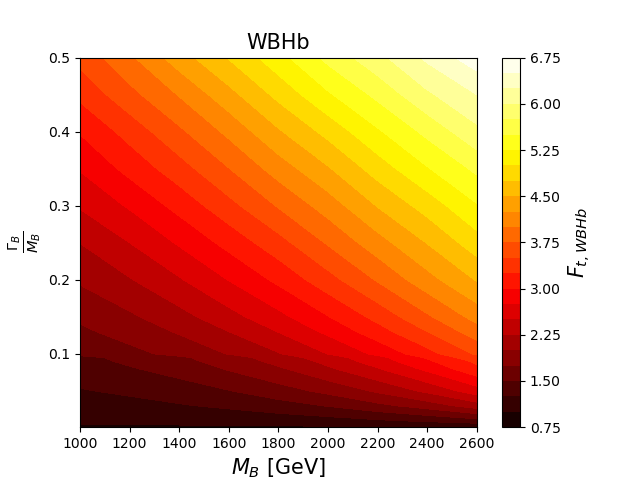}
  \label{fig:Ft-WBHb}
  }
  \caption[]{Calculated values of $F_{VQAq, t}(M_Q, \frac{\Gamma_Q}{M_Q})$ for 
  \protect\subref{fig:Ft-ZTWb} $ZTWb$,
  \protect\subref{fig:Ft-ZTZt} $ZTZt$,
  \protect\subref{fig:Ft-ZTHt} $ZTHt$,
  \protect\subref{fig:Ft-ZBZb} $ZBZb$,
  \protect\subref{fig:Ft-ZBHb} $ZBHb$,
  \protect\subref{fig:Ft-WBWt} $WBWt$,
  \protect\subref{fig:Ft-WBZb} $WBZb$, and
  \protect\subref{fig:Ft-WBHb} $WBHb$ processes.
  } 
  \label{fig:Ftmaps}
\end{figure*}

Following the arguments presented in the previous section and also in Ref.~\cite{Panizzi-LW-3}, we consider the effect of $t$-channel contribution of $WTWb, WTZt, WTHt,$ and $ZBWt$ processes to be marginal and use $F_{t,VQAq} \approx 1$ for these processes. For the processes where the $t$-channel contribution is large, we use the cross-section estimate from the \mbox{\textsc{MadGraph\_aMC@NLO}} simulations described in Section~\ref{sec:topo}. Figure~\ref{fig:Ftmaps} shows the distribution of the $F_{t,VQAq}$ factor as a function of VLQ mass and relative decay widths. It can be seen that there appears to be a lack of continuity, or cutoff, at $\frac{\Gamma}{M} = 0.1$ in these simulations. This cutoff is an artifact of the choice of dynamic scale for the cross-section calculations, determined by \textsc{MadGraph\_aMC@NLO}. The default dynamic scale for these simulations was chosen based on the CKKW scheme~\cite{CKKW-1,CKKW-2}. The \textsc{MadGraph} implementation of this scale choice can vary when particle records with $\frac{\Gamma}{M} > 10\%$ are present in the process which can appear as an scale-dependent uncertainty in the estimation of cross-sections
\footnote{
This observation has been discussed with \textsc{MadGraph} authors and can be followed at \href{https://bugs.launchpad.net/mg5amcnlo/+bug/1943581}{https://bugs.launchpad.net/mg5amcnlo/+bug/1943581}.}. This unphysical,  computational effect introduces an uncertainty of $\mathcal{O}(10\%)$ at the relative decay width threshold of $10\%$.

To obtain parameterized representation of $F_{t, VQAq}$, we independently perform a linear least squared error fit for $\frac{\Gamma}{M} < 0.1$ and a second order polynomial fit for $\frac{\Gamma}{M} \ge 0.1$  for each VLQ mass point simulated (Eqn.~\ref{eqn:Ft-parameter}).

\begin{equation}
    F_{t,VQAq}\left(M_Q, \frac{\Gamma_Q}{M_Q} \right)= 
    \begin{cases}
A_l + B_l\frac{\Gamma_Q}{M_Q}, &  \frac{\Gamma_Q}{M_Q} < 0.1 \\
A_h + B_h\frac{\Gamma_Q}{M_Q} + C_h(\frac{\Gamma_Q}{M_Q})^2, &  \frac{\Gamma_Q}{M_Q} \ge 0.1
\end{cases}
\label{eqn:Ft-parameter}
\end{equation}

\noindent The difference between the fitted polynomials at $\frac{\Gamma}{M} = 0.1$ is treated as an uncertainty associated with the scale dependence of the cross-section estimate for $VQAq$ processes. The best fit values for the parameters in Eqn.~\ref{eqn:Ft-parameter} are given in Table~\ref{tab:Ft-params-low} and Table~\ref{tab:Ft-params-high} for $\frac{\Gamma}{M} < 0.1$ and $\frac{\Gamma}{M} \ge 0.1$ respectively. 

\begin{table*}
\centering
\begin{tabular}{|c|c|c|c|c|c|c|c|c|c|}
\hline
 & $M_{T/B}$ [TeV] &  $ZTWb$  &  $ZTZt$  &  $ZTHt$  &  $ZBZb$  &  $ZBHb$  &  $WBWt$  &  $WBZb$  &  $WBHb$ \\ 
 \hline & 1.0  & 0.984 & 0.969 & 0.975 & 0.973 & 0.981 & 0.977 & 0.982 & 0.981\\ 
 & 1.2  & 0.972 & 0.959 & 0.965 & 0.975 & 0.987 & 0.969 & 0.970 & 0.978\\ 
 & 1.4  & 0.966 & 0.954 & 0.962 & 0.974 & 0.991 & 0.960 & 0.966 & 0.979\\ 
 & 1.6  & 0.959 & 0.951 & 0.963 & 0.977 & 1.002 & 0.956 & 0.963 & 0.985\\ 
$A_l$ & 1.8  & 0.958 & 0.948 & 0.966 & 0.976 & 1.021 & 0.954 & 0.957 & 0.997\\ 
 & 2.0  & 0.951 & 0.945 & 0.972 & 0.979 & 1.043 & 0.950 & 0.952 & 1.017\\ 
 & 2.2  & 0.950 & 0.944 & 0.984 & 0.980 & 1.077 & 0.949 & 0.949 & 1.058\\ 
 & 2.4  & 0.944 & 0.937 & 1.002 & 0.978 & 1.118 & 0.946 & 0.942 & 1.108\\ 
 & 2.6  & 0.947 & 0.938 & 1.032 & 0.977 & 1.176 & 0.944 & 0.944 & 1.195\\ 
\hline 
 & 1.0  & 1.091 & -0.748 & 2.204 & -0.275 & 3.237 & 0.136 & 1.045 & 4.920\\ 
 & 1.2  & 0.959 & -0.471 & 2.807 & -0.103 & 4.196 & 0.084 & 0.983 & 6.206\\ 
 & 1.4  & 0.860 & -0.373 & 3.391 & -0.029 & 5.335 & 0.113 & 0.853 & 7.746\\ 
 & 1.6  & 0.849 & -0.219 & 4.117 & 0.046 & 6.517 & 0.105 & 0.767 & 9.714\\ 
$B_l$ & 1.8  & 0.799 & -0.108 & 4.958 & 0.178 & 7.809 & 0.076 & 0.805 & 12.063\\ 
 & 2.0  & 0.819 & 0.029 & 5.932 & 0.240 & 9.281 & 0.118 & 0.843 & 14.751\\ 
 & 2.2  & 0.849 & 0.071 & 7.057 & 0.348 & 10.727 & 0.110 & 0.810 & 17.689\\ 
 & 2.4  & 0.925 & 0.332 & 8.243 & 0.490 & 12.249 & 0.098 & 0.896 & 21.054\\ 
 & 2.6  & 0.909 & 0.409 & 9.514 & 0.582 & 13.678 & 0.155 & 0.963 & 24.442\\ 
\hline 
\end{tabular}
\caption{Best fit values of the parameters in estimating $F_t$ from Eqn.~\ref{eqn:Ft-parameter} for $\frac{\Gamma}{M} < 0.1$}
\label{tab:Ft-params-low}
\end{table*}

\begin{table*}
\centering
\begin{tabular}{|c|c|c|c|c|c|c|c|c|c|}
\hline
 & $M_{T/B}$ [TeV] &  $ZTWb$  &  $ZTZt$  &  $ZTHt$  &  $ZBZb$  &  $ZBHb$  &  $WBWt$  &  $WBZb$  &  $WBHb$ \\ 
 \hline & 1.0  & 1.070 & 1.068 & 1.122 & 1.067 & 1.147 & 1.035 & 1.070 & 1.126\\ 
 & 1.2  & 1.055 & 1.078 & 1.145 & 1.082 & 1.206 & 1.030 & 1.050 & 1.162\\ 
 & 1.4  & 1.040 & 1.076 & 1.165 & 1.095 & 1.282 & 1.025 & 1.053 & 1.234\\ 
 & 1.6  & 1.035 & 1.080 & 1.222 & 1.103 & 1.376 & 1.025 & 1.042 & 1.335\\ 
$A_h$ & 1.8  & 1.028 & 1.090 & 1.274 & 1.109 & 1.491 & 1.017 & 1.025 & 1.457\\ 
 & 2.0  & 1.039 & 1.083 & 1.373 & 1.113 & 1.638 & 1.018 & 1.041 & 1.699\\ 
 & 2.2  & 1.029 & 1.098 & 1.471 & 1.130 & 1.828 & 1.015 & 1.030 & 1.923\\ 
 & 2.4  & 1.027 & 1.102 & 1.595 & 1.139 & 2.006 & 1.010 & 1.019 & 2.289\\ 
 & 2.6  & 1.020 & 1.101 & 1.771 & 1.130 & 2.217 & 1.006 & 1.029 & 2.689\\ 
\hline 
 & 1.0  & 1.049 & -0.708 & 2.132 & -0.274 & 2.764 & 0.219 & 1.028 & 4.223\\ 
 & 1.2  & 0.898 & -0.588 & 2.447 & -0.175 & 3.334 & 0.144 & 0.919 & 5.133\\ 
 & 1.4  & 0.835 & -0.384 & 2.926 & -0.089 & 3.812 & 0.130 & 0.743 & 6.038\\ 
 & 1.6  & 0.802 & -0.238 & 3.199 & 0.010 & 4.312 & 0.107 & 0.729 & 7.111\\ 
$B_h$ & 1.8  & 0.771 & -0.175 & 3.638 & 0.125 & 4.626 & 0.126 & 0.775 & 8.236\\ 
 & 2.0  & 0.653 & 0.015 & 3.783 & 0.286 & 4.756 & 0.103 & 0.627 & 8.735\\ 
 & 2.2  & 0.706 & 0.027 & 4.029 & 0.280 & 4.559 & 0.106 & 0.710 & 9.749\\ 
 & 2.4  & 0.735 & 0.134 & 4.117 & 0.357 & 4.461 & 0.128 & 0.821 & 9.685\\ 
 & 2.6  & 0.814 & 0.275 & 3.777 & 0.613 & 4.092 & 0.190 & 0.748 & 9.564\\ 
\hline 
 & 1.0  & 0.428 & 0.367 & -1.318 & 0.056 & -1.623 & -0.087 & 0.487 & 1.354\\ 
 & 1.2  & 0.534 & 0.358 & -1.469 & 0.042 & -2.180 & 0.041 & 0.512 & 0.905\\ 
 & 1.4  & 0.575 & 0.185 & -2.013 & 0.036 & -2.656 & 0.059 & 0.704 & 0.417\\ 
 & 1.6  & 0.575 & 0.105 & -2.207 & 0.019 & -3.315 & 0.085 & 0.676 & -0.576\\ 
$C_h$ & 1.8  & 0.616 & 0.150 & -2.760 & -0.034 & -3.752 & 0.067 & 0.628 & -1.710\\ 
 & 2.0  & 0.855 & -0.031 & -2.931 & -0.181 & -4.005 & 0.114 & 0.898 & -2.081\\ 
 & 2.2  & 0.838 & 0.116 & -3.327 & -0.035 & -3.824 & 0.123 & 0.816 & -3.627\\ 
 & 2.4  & 0.884 & 0.078 & -3.554 & -0.005 & -3.876 & 0.105 & 0.706 & -3.455\\ 
 & 2.6  & 0.884 & -0.012 & -3.074 & -0.318 & -3.565 & -0.002 & 0.981 & -3.511\\ 
\hline 
\end{tabular}
\caption{Best fit values of the parameters in estimating $F_t$ from Eqn.~\ref{eqn:Ft-parameter} for $\frac{\Gamma}{M} \ge 0.1$}
\label{tab:Ft-params-high}
\end{table*}


\section{Conclusion}
This letter has thoroughly investigated the effects of non-resonant diagrams on physical distributions of single Top and Bottom partner production at parton level as well as cross-section estimation. Our investigation suggests that the relative strength of the gluon splitting function in the $s-$ and $t-$channel diagrams plays a decisive role in determining the qualitative nature of the impact of non-resonant diagrams in these processes. We additionally evaluate a parameterized correction factor for the inclusive cross-section estimation of these processes. Such factorizable parameterization can facilitate model-independent interpretation of single VLQ searches allowing a harmonized and consistent platform for combination of such searches at the LHC.
\FloatBarrier
\section*{Acknowledgment}
This work is supported by the U.S. Department of Energy, Office of High Energy Physics under Grant No. DE-SC0007890.
\bibliography{main}
\bibliographystyle{unsrt}
\end{document}